 \documentclass[doublespacing]{elsart}

\usepackage{amssymb}
\usepackage[dvips]{graphicx}

\begin{document}

\begin{frontmatter}

% Title, authors and addresses

% use the thanksref command within \title, \author or \address for footnotes;
% use the corauthref command within \author for corresponding author footnotes;
% use the ead command for the email address,
% and the form \ead[url] for the home page:
% \title{Title\thanksref{label1}}
% \thanks[label1]{}
% \author{Steven K. Morrison\corauthref{cor1}\thanksref{label2}}
% \ead{email address}
% \ead[url]{home page}
% \thanks[label2]{}
% \corauth[cor1]{}
% \address{Address\thanksref{label3}}
% \thanks[label3]{}

\title{Tamm states and nonlinear surface modes in photonic crystals}

% use optional labels to link authors explicitly to addresses:
\author{Steven K. Morrison\corauthref{cor1}\thanksref{label1}}
\corauth[cor1]{Tel.: +61 261258277; Fax.:+61 261258588}\ead{skm124@rsphysse.anu.edu.au}
\author[label1]{and Yuri S. Kivshar}
\address[label1]{Nonlinear Physics Center and Center for Ultra-high
bandwidth Devices for Optical Systems (CUDOS), Research School of
Physical Sciences and Engineering, Australian National University,
Canberra ACT 0200, Australia}

\begin{abstract}
We predict the existence of surface gap modes, known as Tamm states
for electronic systems, in truncated photonic crystals formed by two
types of dielectric rods. We investigate the energy threshold,
dispersion, and modal symmetries of the surface modes, and also
demonstrate the existence and tunability of {\em nonlinear Tamm
states} in binary photonic crystals with nonlinear response.
\end{abstract}

\end{frontmatter}

% main text
\section{INTRODUCTION}
Surface modes are a special type of wave localized at the interface
separating two different media. In periodic systems, the modes
localized at the surfaces are known as Tamm
states~\cite{Tamm_ZPhys_32}, first found as localized electronic
states at the edge of a truncated periodic potential. Surface states
have been studied in different fields of physics, including
optics~\cite{Yeh_APL_78}, where such waves are confined to the
interface between periodic and homogeneous media.

Photonic crystals, artificially fabricated periodic structures with
bandgap spectra~\cite{Joannopoulos_book}, can be used for
controlling the properties of light in different devices including
dielectric mirrors and waveguides. In many such applications,
photonic crystals are finite, and are terminated at surfaces
where electromagnetic waves are significantly affected by the
breaking of the translational invariance by the underlying periodic
structure.

Although the properties of electromagnetic waves in bulk photonic
crystals are well understood, studies of electromagnetic waves
near the surfaces of photonic crystals are relatively limited and,
unlike electronic Tamm states, a truncated photonic crystal does not always
support surface states~\cite{Yi_QELS_00}. Surface states in photonic crystals have only
been shown to exist under appropriate changes to the surface layer,
such as a termination though the surface cell or a change to the
surface geometry or material
properties~\cite{Meade_PRB_91,Robertson_OL_94,Ramos_JOSAB_97}.
Furthermore, such surface waves are known to be significantly
sensitive to the surface
termination~\cite{Robertson_OL_94,Vlasov_OL_04}.

In this Communication, we study surface waves in photonic crystals with
terminated (but not altered) surface structures. For the first
time to our knowledge, we show the existence of strongly localized
surface modes in binary photonic crystals, in a complete analogy
to Tamm states in electronic systems. Our analysis of the
linear properties of the diatomic structure illustrates the
dispersion relations and field localization of the surface states,
and the subsequent influence of variations to the crystal geometry.
Additionally, we study the nonlinear behavior of the surface states
when the dielectric function of the photonic crystal includes a
Kerr style nonlinearity, and discuss the nonlinearity-induced
tunability of nonlinear Tamm surface states.

\section{Model and numerical methods}

We consider the propagation of the TM-polarized waves in a
two-dimensional {\em binary photonic crystal} formed by two square
lattices of larger and smaller dielectric rods, as shown in
Fig.~1(a), where $r_a$ and $r_b$ are the rod radii populating the lattices.
These two elemental lattices merge to form a single lattice of
period $a$, with a unit cell created by a single, large rod at the
center of the cell and a single small rod distributed as four
quadrants at the corners of the cell; thereby forming a unit cell
with two rods. In general, the binary structure reduces the
geometric symmetries of the photonic crystal, thereby lifting the
degeneracies of some of the crystal states~\cite{Anderson_PRL96}.
However, the structure we study here does not possess this capacity
and as such maintains the $C_{4v}$ point group symmetry of the
constituent square lattices.

\begin{figure}
\centerline{\scalebox{0.4}{\includegraphics{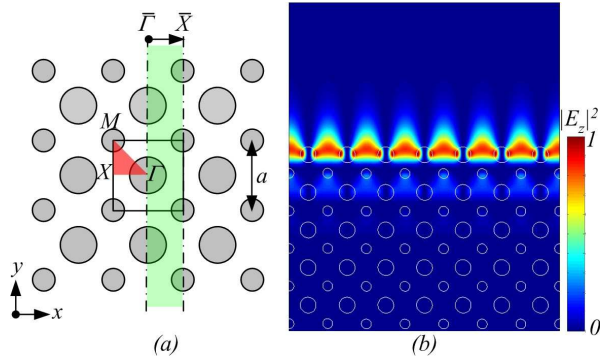}}}
\caption{(a) Structure of a binary photonic crystal with a surface
termination.  Shadings (red and green) show the equivalent irreducible
Brillouin zone of an infinite and semi-infinite photonic crystal
projected onto the crystal geometry. (b) Spatial profile of the
linear surface mode as the color plot of $|E_z|^2$. } \label{model}
\end{figure}

We consider the surface termination that has translational symmetry
along the surface where a row of either large radius rods $r_a$ or
small radius rods $r_b$ form the surface layer exclusively. In all
cases discussed below only complete rods are considered within the
surface layers, similar to the original context of Tamm states.

For these binary crystals, we set the radius of the rods as
$r_a=0.21a$ and $r_b=0.13a$. The dielectric rods of the photonic
crystal correspond to high potential optical regions. In this study,
we set the linear dielectric constant of the rods to
$\varepsilon_{r}=11.56$, corresponding to AlGaAs at a wavelength of
$1550nm$, and assume negligible losses.

To analyze the surface states, we employ two complementary numerical
methods: the plane wave expansion method (PWE)~\cite{Sakoda_book,Johnson_OE00}
and the finite-difference time-domain method (FDTD)~\cite{Taflove_book,Noda_book}.
Due to the normal computational
intensive nature of the FDTD method and the need to reduce the time
step to maintain numerical stability~\cite{Remis_JCP00} for the
nonlinear case, the method is extremely time consuming. To reduce
this burden, we model a semi-infinite crystal using a supercell
representation~\cite{Chan_PRB95,Noda_book} that has a transverse
size of one unit cell, and contains nine unit cells of the photonic
crystal, including the surface cell, and nine unit cells of vacuum
in the lateral direction. The supercell is bordered by a perfectly
matched layer~\cite{Berenger_CP94}. The boundaries
perpendicular to the surface are periodic, and configured with a
$\pi$ spatial phase shift that sets the wave vector to the edge of
the surface Brillouin zone.

\begin{figure}[t]
\centerline{\scalebox{0.4}{\includegraphics{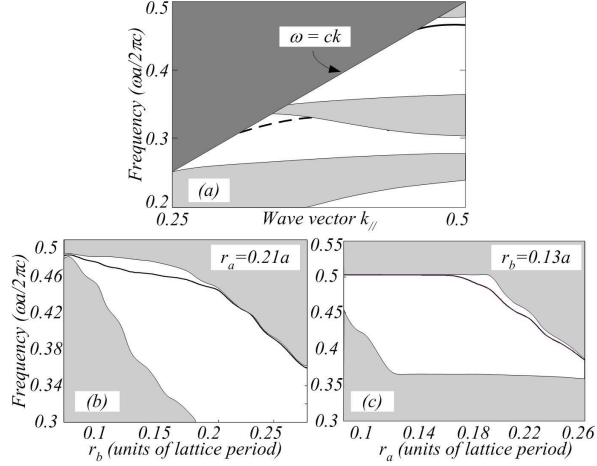}}}
\caption{(a) Surface mode dispersion relationships for the surface formed by larger
(upper solid line) and smaller (lower dashed line) rods. (b,c) Photonic band
gap and surface mode frequencies with respect to dielectric rod radii.
In (b) the larger rods of radius $r_a$ are held constant while the radius $r_b$ of the
smaller rods is varied, whereas in (c) the smaller rods of radius $r_b$ are held constant while the
radius $r_a$ of the larger rods is varied.} \label{sdd}
\end{figure}

The supercell is excited using a two-dimensional Gaussian pulse,
centrally located within the spatial domain and the photonic band
gap, that energizes the resonant modes of the crystal and the
surface structures. A discrete Fourier transform of the time
evolution data, recorded at the surface of the photonic crystal, is
used to determine the spectral response of the surface states. In a
similar manner, the spatial mode profile is calculated from the time
evolution data of the full simulation domain at the peak frequency
of the surface state.

\section{Linear Tamm states}

For the linear surface states, the dispersion relationships are
presented in Fig.~2(a). The dark grey region within the
dispersion diagrams depicts the continuum of free-space states that
exist above the line light ($\omega=ck$) that are not bound to the
surface. Light grey regions within the diagram indicate the Bloch states
of the infinite photonic crystal; states within these regions can
couple into the crystal, and again are not bound to the surface.

Surface states are formed within the second photonic bandgap of the
infinite crystal, and are shown with a solid black line in the
diagram. The surface states occurs within the second band gap when
the surface is terminated in a row of large rods of radius $r_a$. A
localized surface state can form within the first bandgap when
the crystal is terminated in a row of smaller rods, as indicated by
the dashed line in Fig.~2(a). This state, however, can
couple to the Bloch states of the infinite crystal, and we do not
consider it in detail.

We analyze the characteristics of the surface states by varying the
radii of the constituent rods of the photonic crystal.
Figures~2(b,c) show the influence of the rod radii on the
bandgap edges and surface state frequencies. The larger rods
principally define the upper bandgap edge for the $\Gamma-X$
direction and have virtually no effect on the lower bandgap edge,
whereas the smaller rods set the lower bandgap edge with very weak
effect on the upper bandgap edge. Figures 2(b,c) also confirm that the surface mode does not exist when the radii of the two constituent rods are equal

The consistency of the location of the surface states within the
bandgaps signifies the states existence domain is predominantly
defined by the crystal geometry; a feature attributed to the binary
nature of the crystal, which has previously been shown to provide
robust spectral features in the presence of geometric
variations~\cite{Anderson_PRL96}. The spatial field localizations
can be understood by noting the symmetries of the Bloch waves above
and below the bandgap. In the $\Gamma-X$ direction below the second
bandgap, the Bloch wave symmetries place the electric field within
the smaller rods, whereas for above the bandgap the field resides
between the rods in free-space, with some overlap into the larger
rods. As the surface is formed by large rods and the states are near
the upper bandgap edge, they take on the Bloch mode symmetries of
this frequency region.

\section{Nonlinear Tamm states}

We study the properties of nonlinear surface states through the addition of a
third-order susceptibility term within the polarization field of all rods. Throughout our analysis we only modify the
$\chi^{(3)}$ coefficient and maintain the input field intensity and
density. Under steady-state conditions, the principal effect of the
nonlinear term is to induce an intensity dependent change to the
dielectric strength of the surface rods. Simulations of the nonlinear dielectric rods of the photonic crystals are performed using the FDTD
method~\cite{Joseph_IEEE97,Taflove_book}. This is achieved by adding
a $\chi^{(3)}$ term to the polarization as: $ \textbf{P} =
\chi^{(1)}\textbf{E}+\chi^{(3)}|\textbf{E}|^2\textbf{E}$.
The analysis of the linear surface states revealed that the field intensity varies appreciably
across the surface rods. For the nonlinear surface rods this causes
an equivalent change in the dielectric constant. In our analysis of
the nonlinear surface states we consider only a focussing
nonlinearity ( $\chi^{(3)}> 0$) that results in localized increases
in the dielectric index which are proportional to the cube of the
electric field. In turn, this leads to a decrease in the resonant
frequency of the surface state.

Figure~3(a) illustrates the surface mode frequency shift as
the strength of the third-order susceptibility grows. A strong
frequency shift is observed for moderate changes in $\chi^{(3)}$,
with a threshold to the onset of strong frequency shifting at
approximately $\chi^{(3)}=1\times 10^{-3} \mu m^2/V$. These
$\chi^{(3)}$ values exemplify the significant dynamic range of the
nonlinear surface state of a binary photonic crystal using realistic
nonlinearities; for example a typical value of $\chi^{(3)}$ for
AlGaAs is $\chi^{(3)}=8.2 \times 10^{-3} \mu
m^2/V$~\cite{Bahl_PRE03}.

\begin{figure}[t]
\centerline{\scalebox{0.4}{\includegraphics{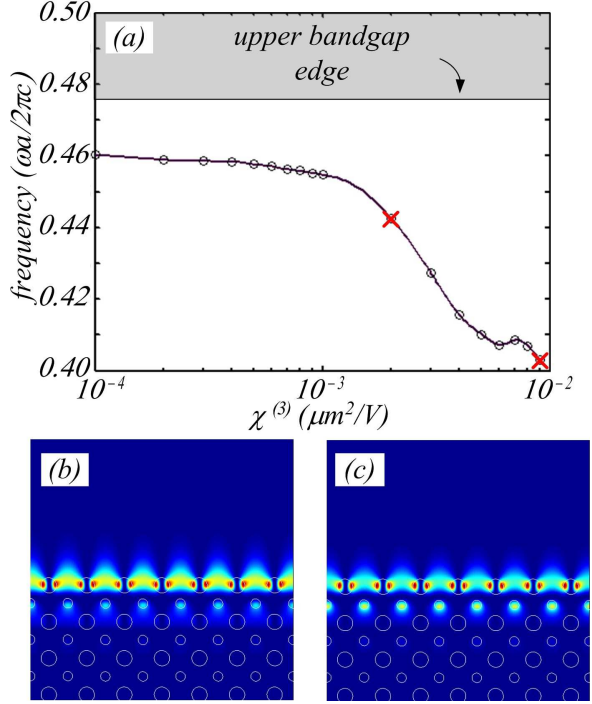}}}
\caption{(a) Nonlinear frequency shift of the surface state vs. the
nonlinear susceptibility for $r_a=0.21a$ and $r_b=0.13a$; (b,c)
Spatial profile of the surface modes as the color plots of the field
density $|E_z|^2$ for $\chi^{(3)}=2\times 10^{-3} \mu m^2/V$ (left),
and $\chi^{(3)}=9\times 10^{-3} \mu m^2/V$ (right).} \label{FvNL}
\end{figure}

The distinct spatial field profile of the surface state, and its
dramatic change with dielectric strength leads to the strong
$\chi^{(3)}$ sensitivity. This is demonstrated in
Fig.~3(b,c), which shows the spatial field profiles for (b) $\chi^{(3)}=2\times 10^{-3} \mu m^2/V$, and (c)
$\chi^{(3)}=9\times 10^{-3} \mu m^2/V$. In a homogeneous nonlinear
focusing medium, a nonlinearity-induced change in the dielectric
constant results in a greater localization of the light, which under
appropriate conditions can lead to \emph{spatial surface solitons}.
However, for the surface state of the binary photonic crystal, the
increased dielectric constant causes the conditions for the surface
resonance to diminish, resulting in a reduction in the field
localization, leading to a complex balance between the nonlinearity
and surface interaction. In general, as the nonlinear surface state
evolves and reaches steady state conditions, the field intensity
within the surface rods reduces. This effect provides optical
limiting that prevents saturation of the nonlinearity and
unrealistic changes to the dielectric strength. Additionally, the
competing effects due to nonlinearity and the surface can lead to
complex behavior, as seen for $\chi^{(3)}=7\times 10^{-3} \mu m^2/V$
where the frequency of the surface state increases. Another
consequence of the nonlinear surface structure is the formation of
a near-surface defect state, as seen in Fig.~3(c), where the
localized state forms in the neighboring row of small rods.

Recent studies have demonstrated the application of optical surface
states of photonic crystals for guiding light, the formation of
high-quality micro cavities, and sub-wavelength imaging~\cite{Yang_APL04,Xiao_arxiv05,Chengyu_JOSAB06}. We
expect that the nonlinear tunable Tamm states found here would
provide substantial nonlinearity-induced control within these
applications leading to novel all-optical surface devices such as
optical limiters and switches.

\section{Conclusion}

In conclusion, we have predicted the existence of surface Tamm
states in binary two-dimensional photonic crystals. Using nonlinear
surface rods, we have highlighted the dynamic tunability of the surface
states resulting from the unusual spatial distribution of the surface
mode, and its geometric transformation through a competition of the
nonlinearity and surface effects.

\section*{Acknowledgment}
This work was produced with the assistance of the Australian Research Council under the ARC Centres of Excellence program.

\end{document}